\documentclass[a4paper]{jpconf}

\usepackage{hyperref,bm,amsmath,amssymb,a4wide,graphicx,subfigure,wasysym,amsfonts,cite}

\begin{document}

\title{Generation of strong magnetic fields in hybrid and quark stars driven by the electroweak interaction of quarks}

\author{Maxim Dvornikov}

\address{Pushkov Institute of Terrestrial Magnetism, Ionosphere and Radiowave Propagation (IZMIRAN), 108840 Troitsk, Moscow, Russia}

\address{Physics Faculty, National Research Tomsk State University, 36 Lenin Avenue, 634050 Tomsk, Russia}

\ead{maxdvo@izmiran.ru}

\begin{abstract}
We study the generation of strong large scale magnetic fields in compact stars containing degenerate quark matter with unbroken chiral symmetry. The magnetic field growth is owing to the magnetic field
instability driven by the electroweak interaction of quarks.
In this system we predict the enhancement of the seed magnetic
field $10^{12}\,\text{G}$ to the strengths $(10^{14}-10^{15})\,\text{G}$.
In our analysis we use the typical parameters of the quark matter
in the core of a hybrid star or in a quark star. We also
apply of the obtained results to model the generation of magnetic fields
in magnetars.
\end{abstract}

The origin of strong magnetic fields $B\sim10^{15}\,\text{G}$ in magnetars~\cite{MerPonMel15} is a puzzle for modern astrophysics. Despite the existence of multiple models describing the generation of such magnetic fields, which are based on magnetohydrodynamics of stellar plasmas, none of them can satisfactory describe the observational data. Recently, in~\cite{Dvo16c}, we proposed the new approach to generate strong magnetic fields in quark matter owing to the instability of the magnetic field driven by the electroweak interaction of quarks. In the present work we summarize the results of~\cite{Dvo16c} and discuss the applicability of this model for the generation of magnetic fields in magnetars.

Let us consider a dense quark matter consisting of $u$ and $d$ quarks.
The density of this matter is supposed to be high enough for the
chiral symmetry to be restored. In this case we can take that quarks
are effectively massless. Therefore we can decompose the quark wave
functions into left and right chiral components, which evolve independently,
and attribute different chemical potentials $\mu_{q\mathrm{L,R}}$,
where $q=u,d$, for each chiral component.

Generalizing the results of~\cite{DvoSem15a,DvoSem15b}, we
get that, in the external magnetic field $\mathbf{B}$, there is the
induced electric current
\begin{equation}\label{eq:Jind}
  \mathbf{J} = \Pi\mathbf{B},
  \quad
  \Pi = \frac{1}{2\pi^{2}}
  \sum_{q=u,d} e_{q}^{2}
  \left(
    \mu_{5q}+V_{5q}
  \right),
\end{equation}
where $e_{u}=2e/3$ and $e_{d}=-e/3$ are the electric charges of
quarks, $e>0$ is the elementary charge, $\mu_{5q}=\left(\mu_{q\mathrm{R}}-\mu_{q\mathrm{L}}\right)/2$
is the chiral imbalance, $V_{5q}=\left(V_{q\mathrm{L}}-V_{q\mathrm{R}}\right)/2$,
and $V_{q\mathrm{L,R}}$ are the effective potentials of the electroweak
interaction of left and right quarks with background fermions. The
potentials $V_{q\mathrm{L,R}}$ were found in~\cite{Dvo15},
\begin{align}\label{eq:VudLR}
  V_{u\mathrm{L}} = & -\frac{G_{\mathrm{F}}}{\sqrt{2}}n_{d}
  \left(
    1-\frac{8}{3}\xi+\frac{16}{9}\xi^{2}-2|V_{ud}|^{2}
  \right),
  \quad
  V_{u\mathrm{R}}=\frac{G_{\mathrm{F}}}{\sqrt{2}}n_{d}
  \left(
    \frac{4}{3}\xi-\frac{16}{9}\xi^{2}
  \right),
  \nonumber
  \\
  V_{d\mathrm{L}} = & -\frac{G_{\mathrm{F}}}{\sqrt{2}}n_{u}
  \left(
    1-\frac{10}{3}\xi+\frac{16}{9}\xi^{2}-2|V_{ud}|^{2}
  \right),
  \quad
  V_{d\mathrm{R}}=\frac{G_{\mathrm{F}}}{\sqrt{2}}n_{u}
  \left(
    \frac{2}{3}\xi-\frac{16}{9}\xi^{2}
  \right),
\end{align}
where $G_{\mathrm{F}}=1.17\times10^{-5}\,\text{GeV}^{-2}$ is
the Fermi constant, $\xi=\sin^{2}\theta_{\mathrm{W}}=0.23$ is the
Weinberg parameter, $n_{u,d}$ are the number densities of $u$ and
$d$ quarks, and $V_{ud}=0.97$ is the element of the Cabbibo-Kobayashi-Maskawa
matrix. The matter of the star is supposed to be electrically neutral.
Thus we should have $n_{u}=n_{0}/3$ and $n_{d}=2n_{0}/3$, where
$n_{0}=n_{u}+n_{d}$ is the total number density of quarks in the
star.  Basing on equation~\eqref{eq:VudLR} and assuming that $n_{0}=1.8\times10^{38}\thinspace\text{cm}^{-3}$, we
get that $V_{5u}=4.5\thinspace\text{eV}$ and $V_{5d}=2.9\thinspace\text{eV}$.


Using equation~(\ref{eq:Jind}) and the results of~\cite{DvoSem15b},
we can obtain the system of kinetic equations for the spectra of the
density of the magnetic helicity $h(k,t)$ and of the magnetic energy
density $\rho_{\mathrm{B}}(k,t)$, as well as the chiral imbalances
$\mu_{5u}(t)$ and $\mu_{5d}(t)$, in the form,
\begin{align}\label{eq:systgen}
  \frac{\partial h(k,t)}{\partial t}= &
  -\frac{2k^{2}}{\sigma_{\mathrm{cond}}}h(k,t) +
  \frac{8\alpha_{\mathrm{em}}}{\pi\sigma_{\mathrm{cond}}}
  \left\{
    \frac{4}{9}
    \left[
      \mu_{5u}(t)+V_{5u}
    \right] +
    \frac{1}{9}
    \left[
      \mu_{5d}(t)+V_{5d}
    \right]
  \right\}
  \rho_{\mathrm{B}}(k,t),
  \nonumber
  \displaybreak[1]
  \\
  \frac{\partial\rho_{\mathrm{B}}(k,t)}{\partial t}= &
  -\frac{2k^{2}}{\sigma_{\mathrm{cond}}}\rho_{\mathrm{B}}(k,t) +
  \frac{2\alpha_{\mathrm{em}}}{\pi\sigma_{\mathrm{cond}}}
  \left\{
    \frac{4}{9}
    \left[
      \mu_{5u}(t)+V_{5u}
    \right] +
    \frac{1}{9}
    \left[
      \mu_{5d}(t)+V_{5d}
    \right]
  \right\}
  k^{2}h(k,t),
  \nonumber
  \displaybreak[1]
  \\
  \frac{\mathrm{d}\mu_{5u}(t)}{\mathrm{d}t}= & 
  \frac{2\pi\alpha_{\mathrm{em}}}{\mu_{u}^{2}\sigma_{\mathrm{cond}}}
  \frac{4}{9}
  \int\mathrm{d}k
  \bigg[
    k^{2}h(k,t)
    \notag
    \\
    & -
    \frac{4\alpha_{\mathrm{em}}}{\pi}
    \left\{
      \frac{4}{9}
      \left[
        \mu_{5u}(t)+V_{5u}
      \right] +
      \frac{1}{9}
      \left[
        \mu_{5d}(t)+V_{5d}
      \right]
    \right\}
    \rho_{\mathrm{B}}(k,t)
  \bigg] -
  \Gamma_{u}\mu_{5u}(t),
  \nonumber
  \displaybreak[1]
  \\
  \frac{\mathrm{d}\mu_{5d}(t)}{\mathrm{d}t} = & 
  \frac{2\pi\alpha_{\mathrm{em}}}{\mu_{d}^{2}\sigma_{\mathrm{cond}}}
  \frac{1}{9}
  \int\mathrm{d}k
  \bigg[
    k^{2}h(k,t)
    \notag
    \\
    & -
    \frac{4\alpha_{\mathrm{em}}}{\pi}
    \left\{
      \frac{4}{9}
      \left[
        \mu_{5u}(t)+V_{5u}
      \right] +
      \frac{1}{9}
      \left[
        \mu_{5d}(t)+V_{5d}
      \right]
    \right\}
    \rho_{\mathrm{B}}(k,t)
  \bigg] -
  \Gamma_{d}\mu_{5d}(t),
\end{align}
where $\Gamma_{u} = 2.98\times10^{-10}\mu_{0}$ and $\Gamma_{d} = 5.88\times10^{-12}\mu_{0}$ are the rates for the helicity flip in $ud$
plasma~\cite{Dvo16c}, $\alpha_{\mathrm{em}}=e^{2}/4\pi=7.3\times10^{-3}$ is the
QED fine structure constant, $\sigma_{\mathrm{cond}}$ is the electric
conductivity of $ud$ quark matter, $\mu_{u}=0.69\mu_{0}$
and $\mu_{d}=0.87\mu_{0}$ are the mean chemical potentials of $u$ and $d$ quarks in the electroneutral
$ud$ plasma, and $\mu_{0}=346\,\text{MeV}$. The wave number $k$ in equation~\eqref{eq:systgen} is in the range:
$k_{\mathrm{min}}<k<k_{\mathrm{max}}$, where $k_{\mathrm{min}}=1/R=2\times10^{-11}\,\text{eV}$,
$R=10\,\text{km}$ is the stellar radius, $k_{\mathrm{max}}=1/\Lambda_{\mathrm{B}}^{(\mathrm{min})}$,
and $\Lambda_{\mathrm{B}}^{(\mathrm{min})}$ is the minimal scale
of the magnetic field, which is a free parameter.

In our model for the magnetic field generation in magnetars, we suggest
that background fermions are degenerate. Nevertheless there is a nonzero
temperature $T$ of the quark matter, which is much less than the
chemical potentials: $T\ll\mu_{q}$. The conductivity of the degenerate
quark matter was estimated in~\cite{HeiPet93}. It can be rewritten in the form~\cite{Dvo16c},
\begin{equation}\label{eq:sigmaT}
  \sigma_{\mathrm{cond}} = \sigma_{0}\frac{T_{0}^{5/3}}{T^{5/3}},
  \quad
  \sigma_{0}=3.15\times10^{22}\,\text{s}^{-1},
\end{equation}
where $T_{0} = (10^{8}-10^{9})\,\text{K}$
is the initial temperature corresponding to the time $t_{0}\sim 10^{2}\,\text{yr}$,
when the star is already in a thermal equilibrium. To derive equation~\eqref{eq:sigmaT} we take that the QCD fine structure constant $\alpha_{s}\sim0.1$. Note that $\sigma_{\mathrm{cond}}$
in quark matter is several orders of magnitude less than the conductivity
of electrons in the nuclear matter in a neutron star (NS)~\cite{Kel73}. Basing on the energy conservation in the system consisting of background fermions and the magnetic field as well as accounting for equation~\eqref{eq:sigmaT}, one gets that the factor
\begin{equation}\label{eq:quenchingF}
  F =
  \left(
    1-\frac{B^{2}}{B_{\mathrm{eq}}^{2}}
  \right)^{5/6},
  \quad
  B_{\mathrm{eq}}^{2} =
  1.23\mu_{0}^{2}T_{0}^{2}.
\end{equation}
should be introduced in rhs of  equation~\eqref{eq:systgen}. It should be noted that the quenching in equation~\eqref{eq:quenchingF} allows one to avoid the excessive growth of the magnetic field when $B \to B_{\mathrm{eq}}$.

While solving of equation~(\ref{eq:systgen}) numerically, we use the initial
Kolmogorov spectrum of the magnetic energy density, $\rho_{\mathrm{B}}(k,t_{0})=\mathcal{C}k^{-5/3}$,
where the constant $\mathcal{C}$ depends on the seed magnetic field $B_{0}$~\cite{DvoSem15b}. The initial spectrum
of the magnetic helicity density is $h(k,t_{0})=2r\rho_{\mathrm{B}}(k,t_{0})/k$,
where the parameter $0\leq r\leq1$, corresponds to initially nonhelical,
$r=0$, and maximally helical, $r=1$, fields.
In our simulations
we shall take that $\mu_{5u}(t_{0})=\mu_{5d}(t_{0})=1\,\text{MeV}$. These initial conditions are quite possible in a dense quark matter
in a hybrid star (HS) or in a quark star (QS)~\cite{Gle00}.

\begin{figure}
  \centering
  \subfigure[]
  {\label{1a}
  \includegraphics[scale=.23]{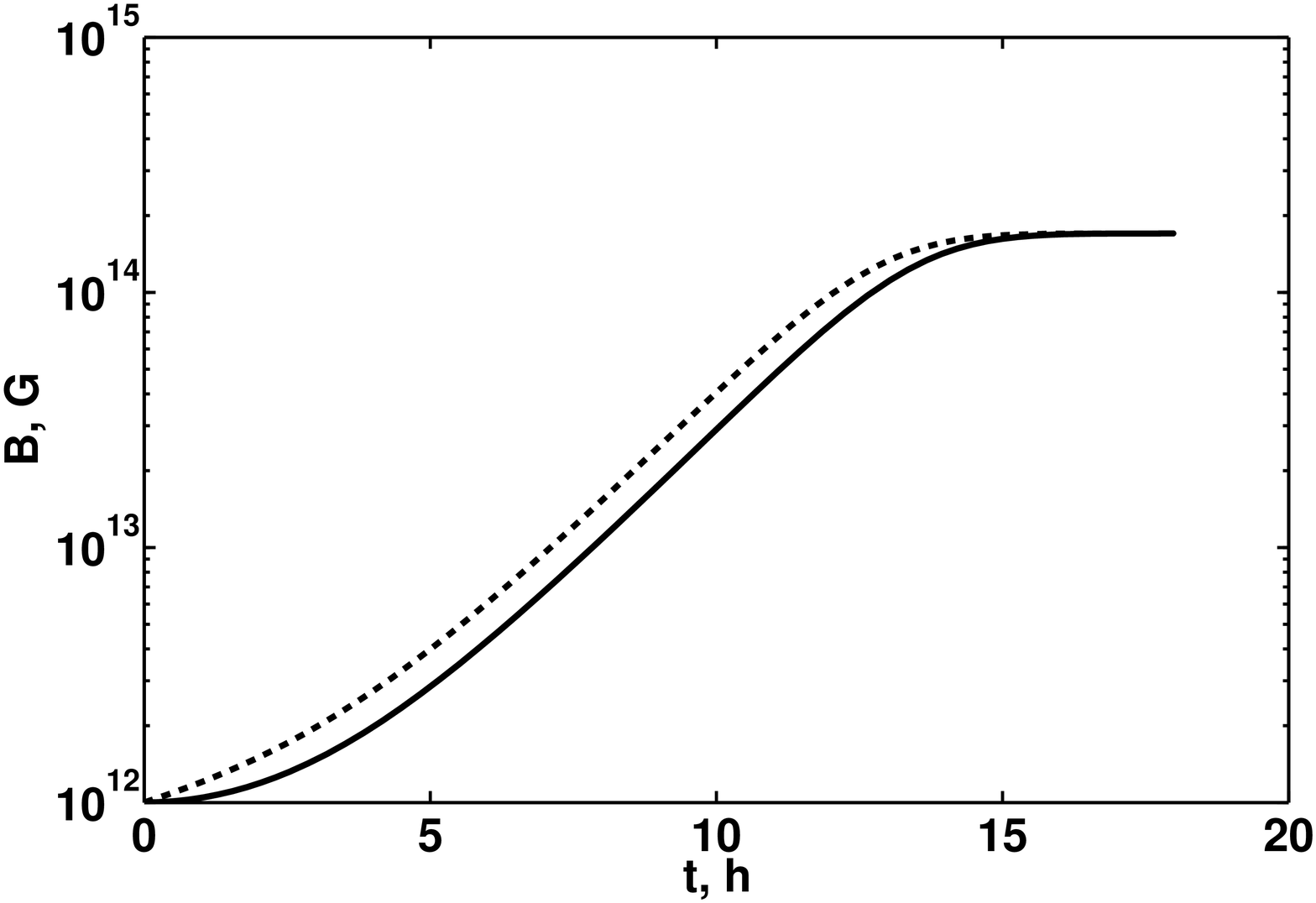}}
  \hskip-.7cm
  \subfigure[]
  {\label{1b}
  \includegraphics[scale=.23]{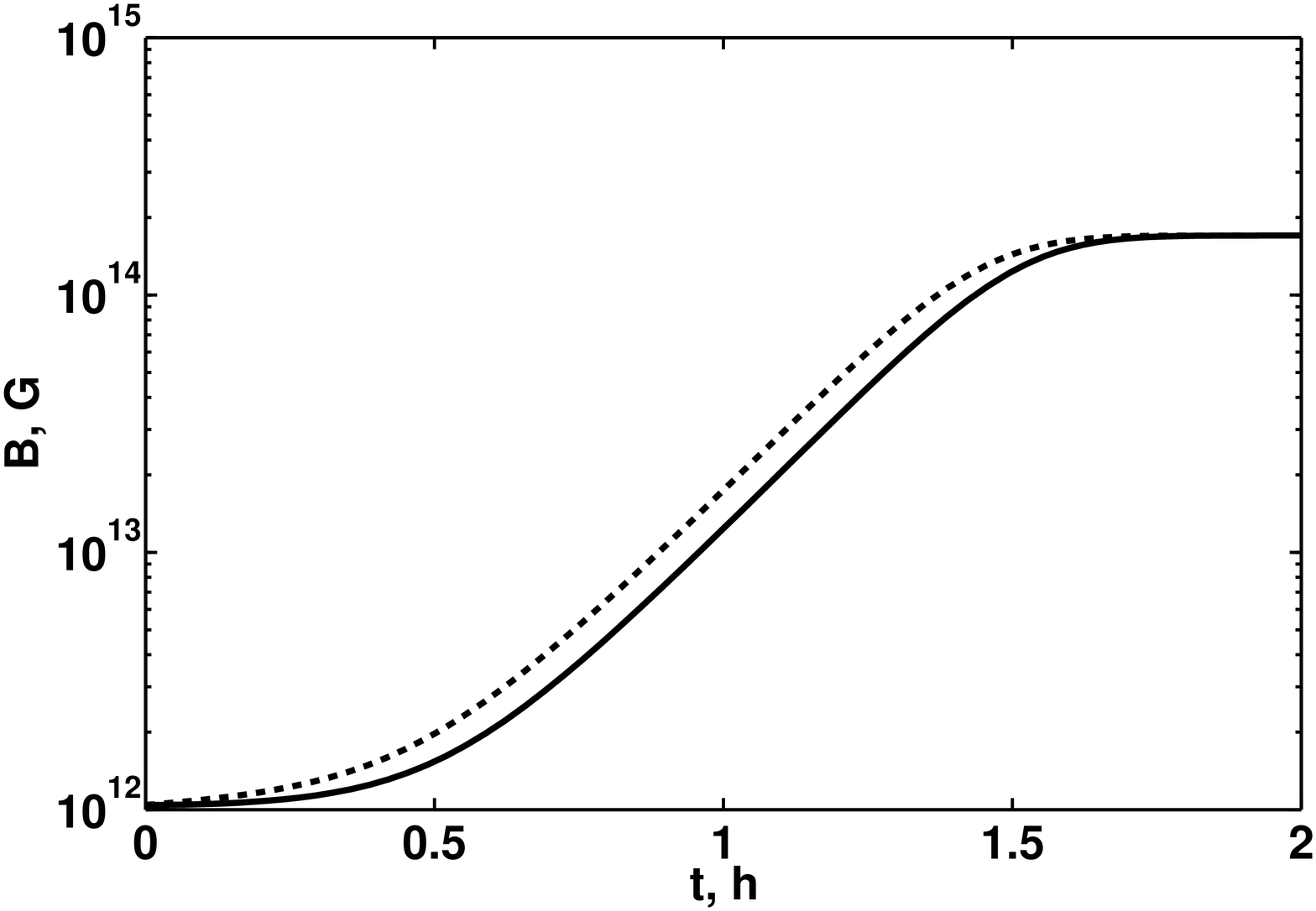}}
  \\
  \subfigure[]
  {\label{1c}
  \includegraphics[scale=.23]{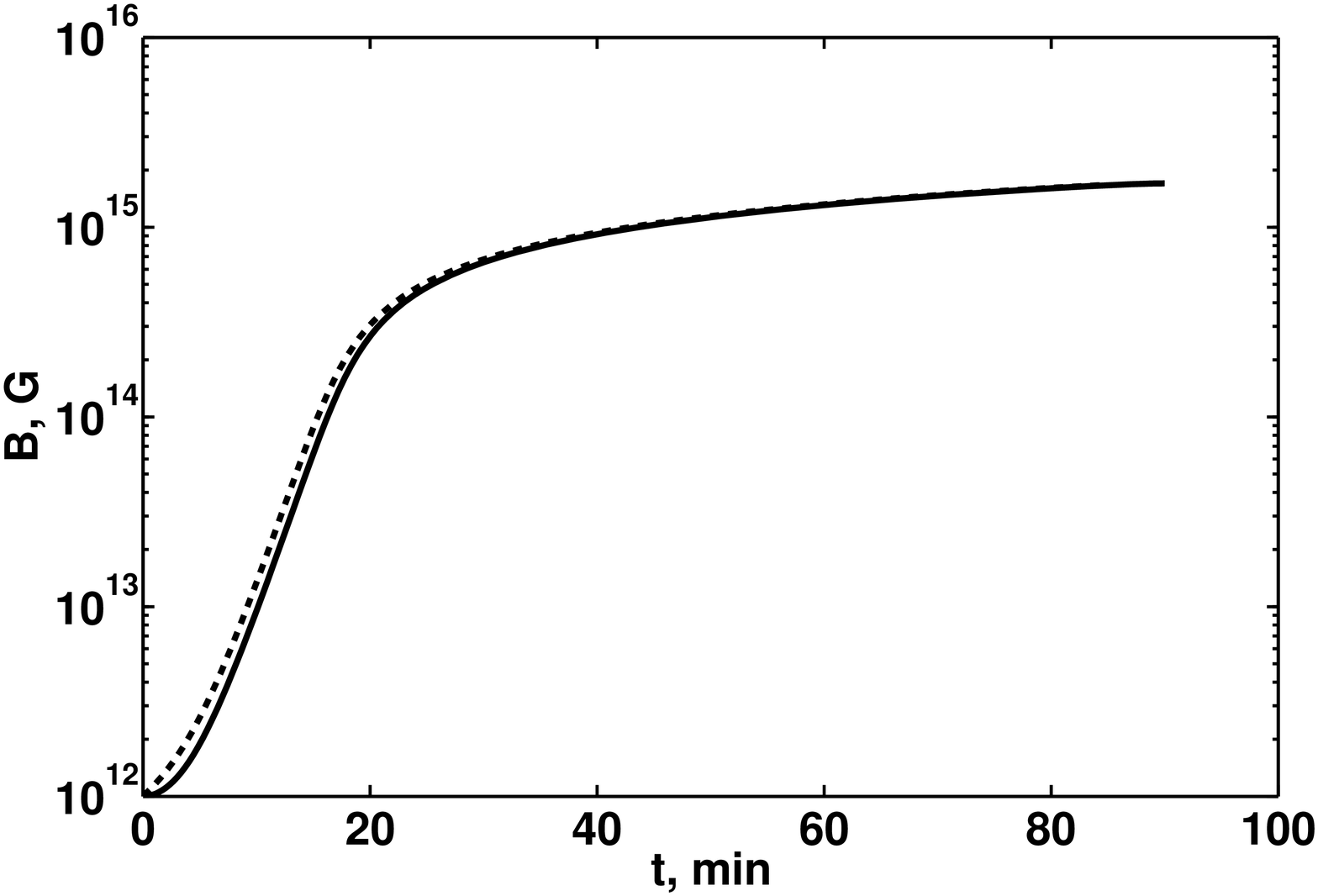}}
  \hskip-.7cm
  \subfigure[]
  {\label{1d}
  \includegraphics[scale=.23]{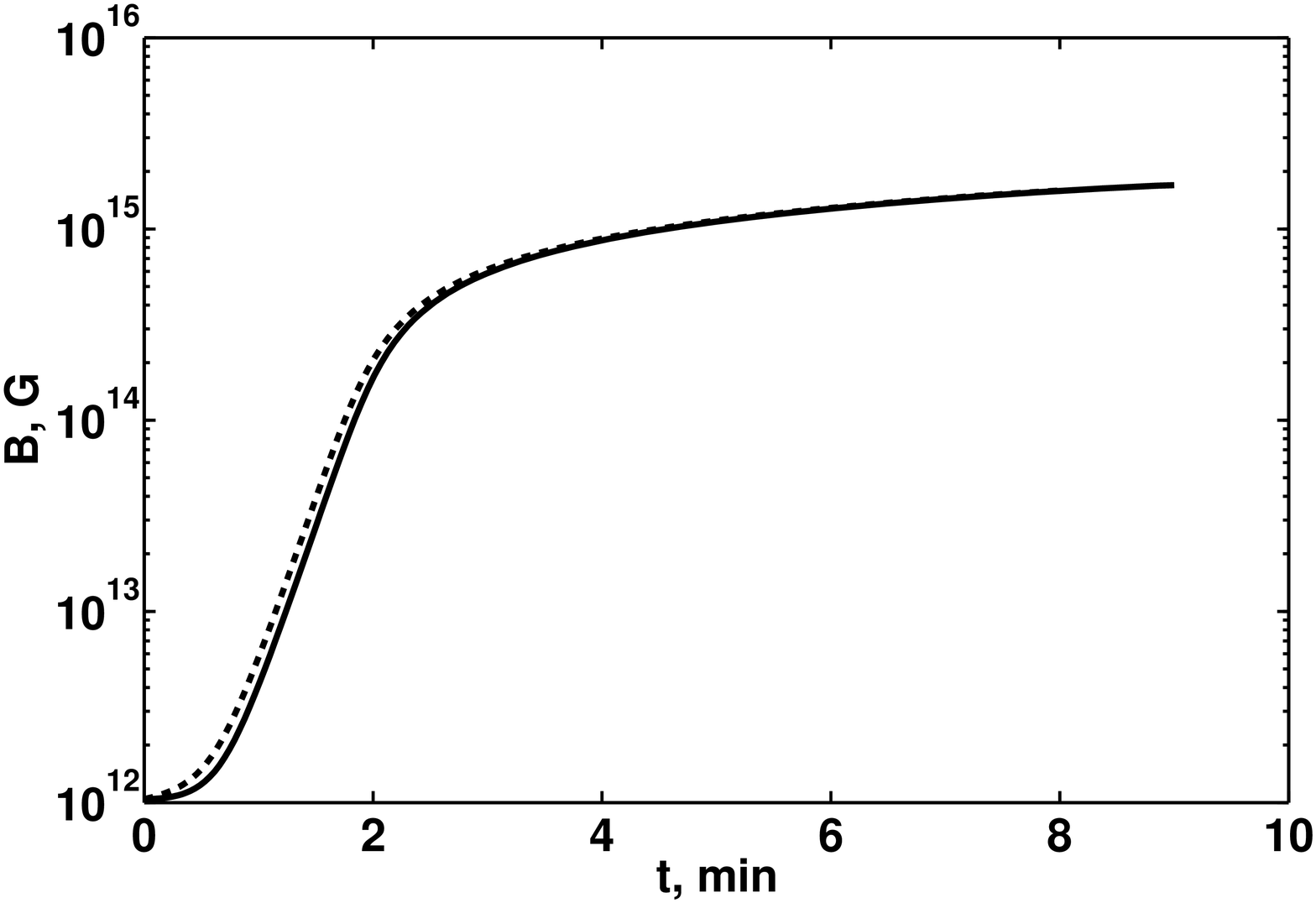}}
  \protect
  \caption{\label{fig:Bevol}
    The magnetic field versus time for different initial temperatures
    $T_{0}$ and minimal length scales   
    $\Lambda_{\mathrm{B}}^{(\mathrm{min})}$.
    The solid lines correspond to initially nonhelical fields with $r=0$
    and dashed ones to the fields having maximal initial helicity, $r=1$.
    (a)~$T_{0}=10^{8}\,\text{K}$ and   
    $\Lambda_{\mathrm{B}}^{(\mathrm{min})}=1\,\text{km}$.
    (b)~$T_{0}=10^{8}\,\text{K}$ and   
    $\Lambda_{\mathrm{B}}^{(\mathrm{min})}=100\,\text{m}$.
    (c)~$T_{0}=10^{9}\,\text{K}$ and 
    $\Lambda_{\mathrm{B}}^{(\mathrm{min})}=1\,\text{km}$.
    (d)~$T_{0}=10^{9}\,\text{K}$ and 
    $\Lambda_{\mathrm{B}}^{(\mathrm{min})}=100\,\text{m}$. 
  }
\end{figure}

In figure~\ref{fig:Bevol} we show the amplification of the initial magnetic
field $B_{0}=10^{12}\,\text{G}$ by two or three orders of magnitude. One can see in figure~\ref{fig:Bevol} that the magnetic field reaches the saturated strength $B_{\mathrm{sat}}$.
This result is analogous to the findings of~\cite{DvoSem15c,Dvo16,Dvo16d}. For $T_{0}=10^{8}\,\text{K}$
in figures~\ref{1a}
and~\ref{1b}, $B_{\mathrm{sat}}\approx1.1\times10^{14}\,\text{G}$;
and for $T_{0}=10^{9}\,\text{K}$ in figures~\ref{1c} and~\ref{1d},
$B_{\mathrm{sat}}\approx1.1\times10^{15}\,\text{G}$. However, unlike~\cite{DvoSem15c,Dvo16,Dvo16d},
$B_{\mathrm{sat}}$ in figure~\ref{fig:Bevol} is defined entirely
by $T_{0}$. The obtained $B_{\mathrm{sat}}$ is close to the magnetic
field strength predicted in magnetars~\cite{MerPonMel15}, especially
if $T_{0}=10^{9}\,\text{K}$.

The time of the magnetic field growth to $B_{\mathrm{sat}}$ is several
orders of magnitude shorter than in~\cite{DvoSem15c,Dvo16,Dvo16d}. This fact
is due to the smaller value of the electric conductivity $\sigma_{\mathrm{cond}}$
in quark matter in equation~(\ref{eq:sigmaT}) compared to
$\sigma_{\mathrm{cond}}$ for electrons in nuclear matter which we used in~\cite{DvoSem15c,Dvo16,Dvo16d}.
This fact can be understood on the basis of equation~(\ref{eq:systgen}), see also~\cite{Dvo16c}. Moreover, we can see that short
scale magnetic field should reach $B_{\mathrm{sat}}$ faster. The later
fact, which was also established in~\cite{DvoSem15b,DvoSem15c,Dvo16,Dvo16d},
is confirmed by the comparison of figures~\ref{1a}
and~\ref{1b} as well as figures~\ref{1c} and~\ref{1d}.

In our model of the magnetic field generation, the thermal energy
of background fermions is converted to the magnetic energy. One can say that a
star cools down magnetically. The typical values of $t_{\mathrm{sat}}$
are $\apprle10\,\text{h}$ in figures~\ref{1a} and~\ref{1b}
and $\lesssim10^{2}\,\text{min}$ in figures~\ref{1c} and~\ref{1d}.
At such short time scales, other cooling channels, such as that due to the neutrino
emission, do not contribute to the temperature evolution
significantly. Therefore, unlike~\cite{DvoSem15b,DvoSem15c,Dvo16,Dvo16d},
we omit them in the present simulations.

In figure~\ref{fig:Bevol} we can see that, although the initial magnetic
helicity can be different (see solid and dashed lines there), the
subsequent evolution of such magnetic fields is almost indistinguishable,
especially at $t \sim t_{\mathrm{sat}}$. It means that, besides the
generation of a strong magnetic field, we also generate the magnetic
helicity in quark matter. This result is in the agreement with~\cite{DvoSem15b,DvoSem15c,Dvo16,Dvo16d}.

In the present work we have applied the mechanism for the magnetic
field generation, proposed in~\cite{DvoSem15a,DvoSem15b,DvoSem15c},
to create strong large scale magnetic fields in dense quark matter.
This mechanism is based on the magnetic field instability driven by a parity violating electroweak
interaction between particles in
the system. We have established the system of kinetic equations for
the spectra of the magnetic helicity density and the magnetic energy
density, as well as for the chiral imbalances, and have solved it numerically.

Although there is a one-to-one correspondence between the mechanisms
for the magnetic field generation in~\cite{DvoSem15a,DvoSem15b,DvoSem15c,Dvo16,Dvo16d}, where we studied the case of NS,
and in the present work, the scenario described here is likely to
be more realistic. As mentioned in~\cite{Dvo16b} the generation
of the anomalous current in equation~(\ref{eq:Jind}) is impossible for
massive particles. Electrons in NS are ultrarelativistic
but have a nonzero mass. The chiral
symmetry can be restored only at densities $n\sim M_{\mathrm{W}}^{3}\sim10^{46}\,\text{cm}^{-3}$,
that is much higher than one can expect in NS. Therefore
the chiral magnetic effect for electrons as well as the results of
Refs.~\cite{DvoSem15a,DvoSem15b,DvoSem15c,Dvo16,Dvo16d} are unlikely to
be applied in NS. Recently this fact was also mentioned in~\cite{Dvo16b}.

On the contrary, the chiral symmetry was found in~\cite{DexSch10}
to be restored for lightest $u$ and $d$ quarks even at densities
corresponding to a core of HS or in QS. Accounting
for the existence of the electroweak parity violating interaction
between $u$ and $d$ quarks, we can conclude that the application
of the methods of~\cite{DvoSem15a,DvoSem15b,DvoSem15c,Dvo16,Dvo16d}
to the quark matter in a compact star is quite plausible.

We have obtained that, in quark matter, the seed magnetic field
$B_{0}=10^{12}\,\text{G}$, which is typical in
a young pulsar, is amplified up to $B_{\mathrm{sat}}\sim\left(10^{14}-10^{15}\right)\,\text{G}$,
depending on the initial temperature.
Such magnetic fields are predicted
in magnetars~\cite{MerPonMel15}. Therefore HS/QS can
become a magnetar. The obtained growth time of the magnetic field
to $B_{\mathrm{sat}}$ is much less than that in electron-nucleon
case studied in~\cite{DvoSem15a,DvoSem15b,DvoSem15c,Dvo16,Dvo16d}.
It means that, in our model, strong magnetic fields are generated
quite rapidly with $t_{\mathrm{sat}}\sim$~several hours after a
star is in a thermal equilibrium.

Note that, in the present work, instead of the quenching of the parameter
$\Pi$ in equation~(\ref{eq:Jind}) suggested in~\cite{DvoSem15c}
to avoid the excessive growth of the magnetic field, we used the conservation of the total
energy and the dependence
of the electric conductivity on the temperature in equation~(\ref{eq:sigmaT}); cf.~\cite{Dvo16,Dvo16d}.
It results in a more explicit saturation of the magnetic field, see equation~\eqref{eq:quenchingF} and figure~\ref{fig:Bevol}.

Summarizing, we have described the generation of strong large scale
magnetic fields in dense quark matter driven by the magnetic field
instability owing to the electroweak interaction of quarks. The described
phenomenon may well exist in the core of HS or in QS.
We suggest that the obtained results can have implication to
the problem of magnetars since the generated magnetic fields have
strength close to that predicted in these highly magnetized compact stars.

\ack

I am thankful to the organizers of ICPPA-2016
for the invitation, to the Tomsk State University Competitiveness Improvement Program and RFBR
(research project No.~15-02-00293) for partial support.

\end{document}